\documentclass[aps,prl,preprint,amsmath]{revtex4}
\usepackage{graphicx}
\usepackage{graphics}
\usepackage{amssymb}
\usepackage{bm}

\begin{document}
\title{Role of Magnetic Plasmon Resonance in Plasmonic Electromagnetically-induced Transparency}


\author{Yuehui Lu,$^1$\ Hua Xu,$^2$\ Nguyen Thanh Tung,$^1$\ Joo Yull Rhee,$^3$\ Won Ho Jang,$^4$\ Byoung Seung Ham,$^2$\ and YoungPak Lee$^{1,}$\footnote{Corresponding author:
yplee@hanyang.ac.kr}}

\affiliation{ $^1$Department of Physics and Quantum Photonic Science
Research Center, Hanyang University, Seoul 133-791, Republic of
Korea\\
$^2$Center for Photon Information Processing, and the Graduate
School of Information and Communications, Inha University,
Incheon 402-751, Republic of Korea\\
$^3$Department of Physics,
Sungkyunkwan University, Suwon 440-746, Republic of Korea\\
$^4$Korea Communication Commission Radio Research Laboratory, Seoul
140-848, Republic of Korea}

\begin{abstract}
We find that the magnetic plasmon resonance plays a vital role in
plasmonic electromagnetically-induced transparency (EIT) proposed by
Zhang et al. [\prl \textbf{101}, 047401 (2008)] as well as the
localized surface plasmon polaritons. Based on this picture, the
plasmonic EIT control through single optical field is suggested in a
modified scheme for active plasmonic switching by simply adjusting
the incident angle of the optical field. The tunability of plasmonic
EIT in this scheme is exhibited sufficiently.

%
\end{abstract}

\maketitle

Electromagnetically-induced transparency (EIT) \cite{RMP05Flesch} is
a quantum optical phenomenon to make an absorptive medium
transparent to a resonant probe field owing to the destructive
quantum interference between two pathways induced by a coupling
field. To excite the EIT in a three-level atomic system, the probe
and the coupling light must satisfy two-photon resonance, and the
Rabi frequency of the coupling light must exceed the effective
dephasing rate of medium. Even though the prerequisite of EIT is
normally satisfied to most alkali atoms, it is not for most solids
owing to a weak oscillator strength and a large dephasing rate.
Although the requirement of EIT such as sharp-energy-level
transitions with dipole-forbidden ground states looks somewhat
critical in application viewpoint, the interest in EIT has been
grown up because of the potential applications of optical switching
using EIT-enhanced giant Kerr nonlinearity \cite{OL96Schmidt} in
both nonslow-light \cite{PRL00Ham} and slow-light regimes
\cite{PRL99Harris}. Especially, in an N-type system, where a second
control field is added to the EIT scheme, the ultraweak
control-field-triggered optical switching \cite{PRA08Ham} gives a
great benefit in nanophotonics as well as quantum information.


Unlike the quantum interference in atomic systems, the coupled
components based on different mechanisms can be applied to realize
the analogy of EIT in linear classical systems, such as mechanical
oscillators, \emph{RLC} circuits \cite{AMP02Alzar}, optical
resonators \cite{PRA01Opatrny}, optical dipole antennas
\cite{PRL08SZhang,arXiv09HXu}, trapped-mode patterns
\cite{PRL08Papasimakis}, split-ring resonators (SRRs)
\cite{PRL09Tassin}, and array of metallic nanoparticles
\cite{PRB09Yannopapas}. As a scheme of EIT-like effect at optical
frequencies proposed by Zhang et al. \cite{PRL08SZhang} and Xu et
al. \cite{arXiv09HXu}, the application of optical dipole antennas is
of great interest with the feature of subwavelength, since the
half-wavelength scaling breaks down when the localized surface
plasmon polaritons (SPPs) play a role in the electromagnetic
response \cite{PRL07Novotny}. In the unit cell, the optical dipole
antenna comprising a simple metal strip is termed as the ``bright
mode'' while the other two parallel strips are the ``dark mode,''
depending on how strong an incident light from free space can be
coupled into the plasmonic mode \cite{PRL01Stockman}. It is
generally agreed that the EIT-like effect was ascribed to the
coupling of the bright mode with the dark one, the strength of which
can be controlled by the spatial separation of two plasmonic modes.
However, how the coupling works in the EIT-like effect still remains
unknown.

In this Letter, we analyze the underlying physics of the classical
version of EIT in a metamaterial, which helps us specifically
understand the coupling process of the bright and the dark modes.
Based on it, we propose an active plasmonic switching without using
the coupling/control fields required in the conventional EIT scheme.

To illustrate the physical picture of the coupling between the
bright and the dark modes, we first consider a typical unit cell, as
shown in Fig. \ref{fig1}(a), whose geometrical parameters are taken
from Ref. \cite{PRL08SZhang}. The permittivity of metal,
$\epsilon_m$, is modeled as silver using the Drude formula, where
the plasma frequency and the collision frequency are $\omega_p =
1.366\times10^{16}$ rad/s and $\gamma = 3.07\times10^{13}$ Hz,
respectively. The numerical simulation is carried out using a
finite-integration package, \emph{CST Microwave Studio}.

A single metal strip can be electrically polarized by a plane wave,
working as an optical dipole antenna and providing the absorption
background. In the near zone, the electric dipole fields
\cite{CE99Jackson} approach in the Syst\`{e}me International (SI)
Units
\begin{subequations}\label{eq1}
\begin{equation}
\textbf{H}_1=\frac{i\omega}{4\pi}(\textbf{n}\times\textbf{p})\frac{1}{r^2},
\end{equation}
\begin{equation}
\textbf{E}_1=\frac{1}{4\pi\epsilon_0}[3\textbf{n}
(\textbf{n}\cdot\textbf{p})-\textbf{p}]\frac{1}{r^3},
\end{equation}
\end{subequations}
where $\textbf{n}$ and $\textbf{p}$ are the unit vector in the
direction of propagated field and the electric dipole moment,
respectively. There is a $\pi/2$ phase retardation of the magnetic
component with respect to the electric one in the dipole electric
fields according to Eq. \ref{eq1}, which differs from those in the
radiation zone where they are in phase
\cite{CE99Jackson,OPT02Hecht}.

Generally speaking, a dark mode cannot be excited efficiently by the
external electromagnetic irradiation, since the external electric
and magnetic fields are perpendicular and parallel to the two metal
strips, respectively. However, we note that the magnetic field
$\textbf{H}_1$ from the fields of the dipole optical antenna is
normal to the plane of two parallel metal strips, which might
stimulate the dark mode based on the magnetic plasmon resonance
(MPR) \cite{PRB07Sarychev}. According to Faraday's law,
\begin{equation}\label{eq2}
\nabla \times \textbf{E}_2 =
-\mu_0\mu_d\frac{\partial}{\partial{t}}(\textbf{H}_1 +
\textbf{H}_\textrm{in}),
\end{equation}
where $\textbf{H}_\textrm{in}$ and $\mu_d$ are the magnetic field
induced by the circular current \cite{OL05Shalaev} and the
permeability of dielectric medium, respectively. the second electric
field $\textbf{E}_2$ is excited by the time-dependent magnetic field
$\textbf{H}_1$ and the induced one $\textbf{H}_\textrm{in}$.
Integration of Eq. \ref{eq2} over the dotted-contour $\{1, 2, 3,
4\}$ in Fig. \ref{fig1}(b) yields the following differential
equation \cite{PRB07Sarychev}:
\begin{equation}\label{eq3}
\left[{2}I(y)Z + \frac{\partial{U}}{\partial{y}}\right]\triangle{y}
= -\mu_0\mu_ds\left[\dot{I}(y) + \dot{H}_{1z}(y)\right]\triangle{y},
\end{equation}
where $I(y)$, $Z$ and $U$ are the surface current density, the
surface impedance $Z = i/(\epsilon_0\epsilon_m{W_2}\omega)$
($\epsilon_m$ is the metal permittivity) and the potential drop,
respectively. $\triangle{y}$ is the distance between points 1 and 4.
$\dot{I}(y)$ and $\dot{H}_{1z}(y)$ are the time derivatives. $U(y)$
has the form of
\begin{equation}\label{eq4}
U(y) = sE_{2x} = \frac{s}{\epsilon_0}[Q(y)-P(y)],
\end{equation}
where $Q(y)$ and $P(y)$ are the electric charge in unit area and the
medium polarization. The charge conservation law leads to
\begin{equation}\label{eq5}
\frac{\partial{I}}{\partial{y}} = -\dot{Q}.
\end{equation}
We substitute $U(y)$ into Eq. \ref{eq3} and take the time derivative
on both sides. With the help of Eq. \ref{eq5}, it yields
\begin{equation}\label{eq6}
\frac{d^2I(y)}{dy^2} = -g^2I(y) - tH_{1z}(y),
\end{equation}
where $\epsilon_d$ is the permittivity of dielectric medium.
Parameters $t$ and $g$ is expressed as:
\begin{subequations}\label{eq7}
\begin{equation}
t = \mu_0\mu_d\epsilon_0\epsilon_d\omega^2
\end{equation}
and
\begin{equation}
g^2 = t - \frac{2\epsilon_d}{W_2s\epsilon_m},
\end{equation}
\end{subequations}
respectively. Solving Eq. \ref{eq6}, the induced current, $I(y)$,
can be obtained under the assumption that $H_{1z}(y)$ is independent
on $y$ in a finite area and the initial conditions are $dI(0)/dy =
I(L_2) = 0$. Thus, the induced magnetic and electric fields,
$\textbf{H}_\textrm{in}$ and $\textbf{E}_2$, can be written as
\begin{subequations}\label{eq8}
\begin{equation}
H_{\textrm{in}z}(y) =
\frac{t}{g^2}H_{1z}\left[\frac{\cos(gy)}{\cos(gL_2)} - 1\right],
\end{equation}
\begin{equation}
E_{2x}(y) =
i\frac{t}{\epsilon_0\epsilon_d\omega{g}}H_{1z}\frac{\sin(gy)}{\cos(gL_2)},
\end{equation}
\begin{equation}
E_{2y}(x,y) =
i\omega\mu_0\mu_dxH_{1z}\left[\left(1+\frac{t}{g^2}\right)\frac{\cos(gy)}{\cos(gL_2)}
+ \left(1-\frac{t}{g^2}\right)\right].
\end{equation}
\end{subequations}
Again, a phase retardation of $\pi/2$ appears when the electric
component is compared with the magnetic one.

Taking into account the former $\pi/2$ retardation from Eq.
\ref{eq1}, the total phase retardation comes to be $\pi$ between two
electric fields $\textbf{E}_1$ and $\textbf{E}_2$, which causes
their destructive interference. Consequently, once the localized
electric field $\textbf{E}_1$ around the dipole antenna is partially
suppressed, the transmission of the incident field is expected to be
boosted considerably. This implies that the excitation of MPR
contributes to the transmission within an absorption background
arising from the localized SPPs. Furthermore, since the interference
occurs in the near zone, it is likely to achieve the EIT-like effect
even using a single unit, instead of a periodic array, which can
make the device more compact.

Although the optical destructive interference is totally
distinguished from the quantum interference, there is still an
analogy of the pathway picture in quantum EIT. The dipole optical
antenna is excited by the external electromagnetic irradiation,
which is similar to an atomic excitation from a ground state to a
excited state. Then, the induced magnetic field by the dipole
optical antenna works as a metastable state, which produces the MPR,
as implied by Fig. \ref{fig1}(c), and the resultant electric field
affects the former dipole electric field owing to the phase
difference of $\pi$. Thus, the superposition of these two electric
fields gives rise to the destructive interference as if the
transitions between two different pathways render the quantum
interference, which provides the EIT-like effect with a vivid
classical version.

According to the aforementioned analysis, the MPR plays a decisive
role in the EIT-like feature at optical frequencies, although it has
not been recognized that the localized SPPs and the MPR coexist
\cite{PRL09Tassin}, where the effective permittivity,
$\epsilon_\textrm{{eff}}$, and the effective permeability,
$\mu_\textrm{{eff}}$, are treated separately. The direct way to
confirm this coexistence of the localized SPPs and the MPR is to
retrieve the effective parameters using the extraction approach in
Ref. \cite{PRE04XChen}. As expected, not only the electric plasmon
resonance would be excited, but also the MPR would emerge, as shown
in Fig. \ref{fig2}. In addition, the imaginary parts of
$\epsilon_\textrm{{eff}}$ and $\mu _\textrm{{eff}}$ are opposite in
sign, which does not contradict the fundamental laws in nature
\cite{PRE03Koschny}.

Based on the role of MPR, we present a modified scheme, as shown in
Fig. \ref{fig3}, where one more dipole optical antenna is added to
form a symmetric structure. Equation \ref{eq1} tells that the
induced magnetic fields, $H_{1z}$, from the left dipole antenna in
Fig. \ref{fig3} has the same amplitude, but the opposite direction,
comparing with that from the right one. These induced magnetic
fields lead to the opposite circular currents, which cancel out each
other. Thus, the MPR is suppressed, as illustrated in Fig.
\ref{fig4}(a), where the magnetic field vanishes between two
parallel metal strips in the modified structure, in contrast to that
in the typical structure. This means that the EIT-like feature in
Fig. \ref{fig1}(c) is destroyed owing to the antisymmetric MPRs, as
in Fig. \ref{fig4}(b), like in a resonant N-type four-level atomic
system. The antisymmetric MPRs in Fig. \ref{fig4}(a), however,
changes if an optical field is irradiated at an oblique angle, where
the wavevector, $\textbf{k}$, varies in the $yz$ plane. This is
because the balance between two induced magnetic fields in Fig.
\ref{fig4}(a) is broken by the vertical magnetic component $H_{0z}$
of plane wave, $H_{0z} = H_0\sin\theta$, where $H_0$ and $\theta$
denote the magnetic field of incident field and the incident angle,
respectively. Figure \ref{fig4}(c) shows the resultant effect of
EIT-like feature, where the magnetic field is considerably localized
between the two parallel metal strips like in Fig. \ref{fig1}(c), at
an incident angle of 15$^\circ$. Correspondingly, the plasmonic EIT
is obtained as shown in Fig. \ref{fig4}(d) even with the symmetric
design by controlling the incident angle of the optical field.

In conclusion, we propose an active control of the plasmonic EIT in
a metamaterial based on MPR, which plays a critical role in it. The
generation of MPR, further confirmed by the analytical expressions
and the effective permeability, plays as a control field in an
N-type atomic system. According to the physical picture we present,
the plasmonic EIT is the result of optical destructive interference
and is the classical version of atomic EIT. Furthermore, the bright
and the dark modes can be either coupled or uncoupled, depending on
the angle of incidence in the symmetric structure. Therefore, the
active control of plasmonic EIT can be implemented even using only
an external field by controlling the incident angle.

This work was supported by MEST/NRF through the Quantum Photonic
Science Research Center, Korea.

\newpage
\begin{center}
\textbf{Figure captions}
\end{center}

Fig. 1 (color online). (a) Schematic of a typical unit cell
consisting of a bright element (single metal strip) and a dark
element (two parallel metal strips), whose geometrical parameters
$W_1$, $L_1$, $W_2$, $L_2$, $s $ and $d$ are 50, 128, 30, 100, 30
and 40 nm, respectively. The thickness of all metal strips is 20 nm.
The single metal strip is electrically polarized by the plane wave,
working as an optical dipole antenna. (b) Circular current is
excited by magnetic field $\textbf{H}_1$ from the fields of optical
dipole antenna. (c) Amplitude distribution of the magnetic field at
the plasmonic EIT.

Fig. 2 (color online). (a) Real (black curve) and imaginary (blue
curve) parts of the effective permittivity
$\epsilon_\textrm{{eff}}$, and (b) those of the effective
permeability $\mu_\textrm{{eff}}$.

Fig. 3. Schematic of a unit cell consisting of two bright elements
(metal strips on the left and the right) and a dark element (two
parallel metal strips in the middle), whose geometrical parameters
are the same as those in Fig. \ref{fig1} except $s = 80$ nm.

Fig. 4 (color online). (a) Amplitude distribution of the magnetic
field (at the absorption peak of 471 THz) and (b) transmission
spectrum at normal incidence. (c) Amplitude distribution of the
magnetic field (at the peak of plasmonic EIT of 471 THz) and (d)
transmission spectrum at an incident angle of 15$^\circ$ (the
reflection plane located in the $yz$ plane).

\clearpage
\begin{figure}
\includegraphics{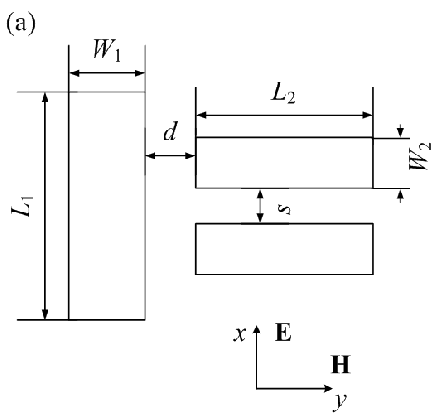}
\end{figure}
\begin{figure}
\includegraphics{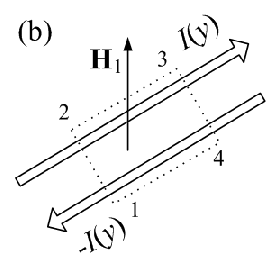}
\end{figure}
\begin{figure}
\includegraphics{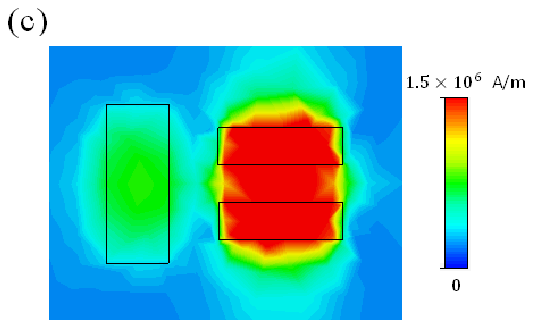}
\caption{Lu et al.}\label{fig1}
\end{figure}

\clearpage
\begin{figure}
\includegraphics{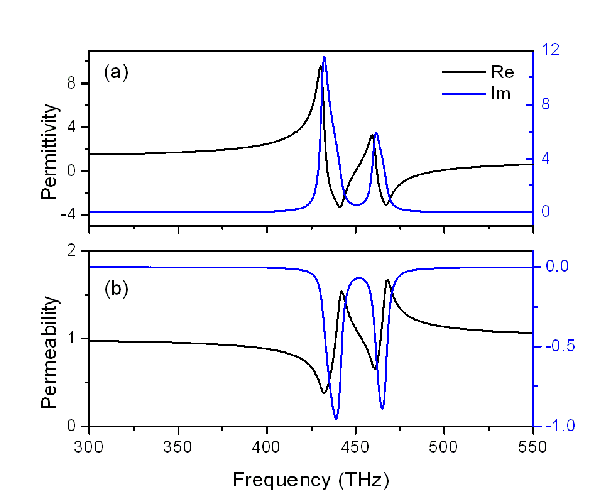}
\caption{Lu et al.} \label{fig2}
\end{figure}

\clearpage
\begin{figure}
\includegraphics{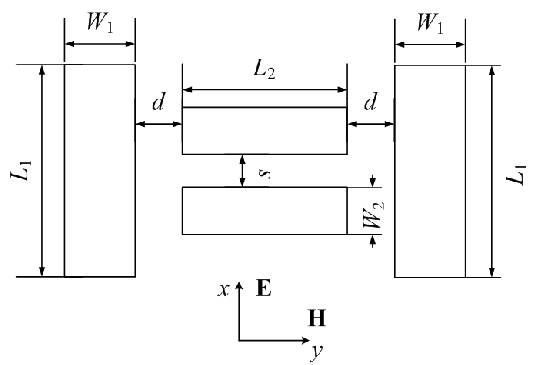}
\caption{Lu et al.} \label{fig3}
\end{figure}

\clearpage
\begin{figure}
\includegraphics{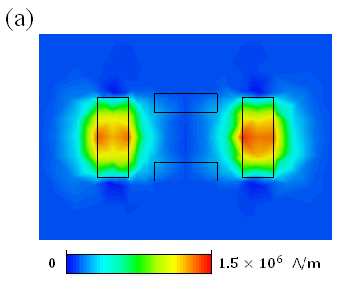}
\end{figure}
\begin{figure}
\includegraphics{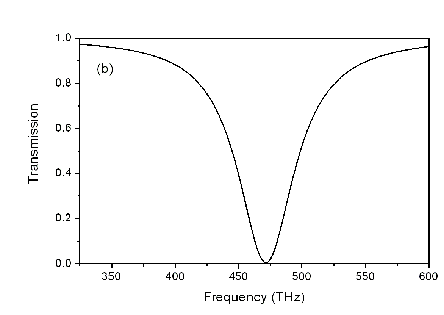}
\end{figure}
\begin{figure}
\includegraphics{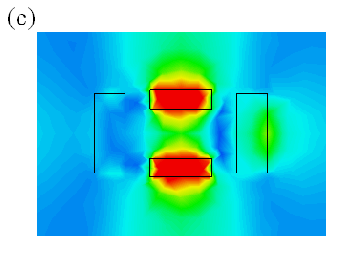}
\end{figure}
\begin{figure}
\includegraphics{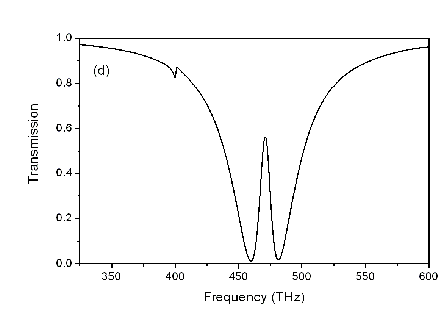}
\caption{Lu et al.}\label{fig4}
\end{figure}

\end{document}